

\input phyzzx        
\input epsf          

\catcode`\@=11 
\def\NEWrefmark#1{\step@ver{{\;#1}}}
\catcode`\@=12 

\overfullrule 0pt    

\baselineskip 13pt plus 1pt minus 1pt
\nopubblock
{~} \hfil \vbox{\hbox{MIT-CTP-2158}\hbox{hep-ph/9307329}\hbox{July 1993}}
\break
\titlepage
\title{NOVEL QUARK FRAGMENTATION FUNCTIONS
AND THE NUCLEON'S TRANSVERSITY DISTRIBUTION}
\author{R. L. Jaffe\foot{Supported in part by funds provided by the U.S.
Department of Energy (D.O.E.) under contract \#DE-AC02-76ER03069,
and in part by the Texas National Research Laboratory Commission under
grants \#RGFY92C6 and \#RGFY93278C\hfill\break
E-mail address:  jaffe@mitlns.mit.edu.}
and Xiangdong Ji
\foot{Supported in part by funds provided by the U.S.
Department of Energy (D.O.E.) under contract
\#DE-AC02-76ER03069.\hfill\break E-mail  address:  xdj@mitlns.mit.edu}}
\address{ Center for Theoretical Physics\break
Laboratory for Nuclear Science and Department of Physics\break
Massachusetts Institute of Technology\break
Cambridge, MA 02139, USA}

\abstract {We define twist-two and twist-three
quark fragmentation functions in Quantum Chromodynamics (QCD)
and study their physical implications.
Using this formalism we show how the nucleon's transversity distribution
can be measured in single pion inclusive electroproduction.}

\endpage

\def\define#1#2\par{\def#1{\Ref#1{#2}\edef#1{\noexpand\refmark{#1}}}}
\def\con#1#2\noc{\let\?=\Ref\let\<=\refmark\let\Ref=\REFS
         \let\refmark=\undefined#1\let\Ref=\REFSCON#2
         \let\Ref=\?\let\refmark=\<\refsend}

\let\refmark=\NEWrefmark

\define\EMC{The EMC Collaboration, J. Ashman {\it et al.\/}, {\it Nucl.
Phys.} {\bf B328} (1989) 1.}

\define\SLAC{The E142 Collaboration, P. L. Anthony {\it et al.\/},
SLAC-PUB-6101 (1993)}

\define\SMC{The SMC Collaboration, B. Adeva {\it et al.\/}, {\it Phys.
Lett.\/} {\bf B302} (1993) 533.}

\define\PennState{J. C. Collins, {\it Nucl. Phys.} {\bf B394} (1993) 169.}

\define\JaffeJi{R. L. Jaffe and X. Ji, {\it Phys. Rev. Lett.\/} {\bf 67}
(1991) 552.}

\define\RS{J. P. Ralston and D. E. Soper, {\it Nucl. Phys.\/} {\bf B152}
(1979) 109.}

\define\Collins{J. C. Collins, S. F. Heppelmann, and G. A. Ladinsky,
Penn. State Preprint PSU/TH/101, (1993).}

\define\Artru{X. Artru and M. Mekhfi, {\it Z. Phys.\/} {\bf C45} (1990)
669.}

\define\CollinsSoper{J. C. Collins and D. Soper, {\it Nucl. Phys.\/} {\bf
B194} (1982) 445.}

\define\Ji{X. Ji, MIT Preprint MIT-CTP-2219, 1993.}

\define\JaffeSoldate{R. L. Jaffe and M. Soldate, in {\it Perturbative
Quantum Chromodynamics\/}, Proceedings of the Tallahassee Conference, 1981,
D. W. Duke and J. F. Owens, eds. (AIP, New Yor, 1981), p. 60; {\it Phys.
Lett.\/} {\bf 105B} (1981) 467;{\it Phys. Rev.\/} {\bf D26} (1982) 49.}

\define\EFP{R. K. Ellis, W. Furmanski and R. Petronzio {\it Nucl. Phys.\/}
{\bf B212} (1983) 29.}

\define\Jaffe{R. L. Jaffe, {\it Comm. Nucl. Part. Phys.} {\bf 14} (1990)
239.}

\define\Rith{M. Veltri, et al., in Hamburg Proceedings, Physics at
   HERA, vol. 1 447-457, 1991. }

\define\ShuryakVainshteyn{E. Shuryak and Vainshteyn, {\it Phys. Lett.\/} {\bf
105B} (1981) 65; {\it
Nucl. Phys.\/} {\bf B199} (1982) 451.}

\FIG\fragfunction{Diagram for quark fragmentation functions.}

\FIG\transversity{Pion-production in deep-inelastic scattering.}



Recent measurements of the nucleon's dominant spin--dependent quark
distribution have sparked renewed interest in
deep--inelastic spin physics.$^{\EMC,\SLAC,\SMC}$~  In addition to the
debate over the unexpected result reported in Ref.~[\EMC], the
classification and interpretation of spin dependent effects in deep
inelastic scattering has been re--examined and
extended.$^{\PennState,\JaffeJi}$~  One of the most interesting consequences
has been the discovery of a class of {\it chirally odd\/} quark
distribution functions including one, $h_1(x,Q^2)$,$^{\JaffeJi,\RS}$ which
scales in the deep inelastic limit and provides the long-sought parton
description of the quark distribution in a transversely polarized
nucleon.$^{\JaffeJi}$~ For reasons discussed in Ref.~[\JaffeJi] we call
$h_1(x,Q^2)$ the nucleon's {\it transversity\/} distribution.
Chirally odd quark distributions are difficult to
measure because they are suppressed in totally--inclusive deep inelastic
scattering.  Up to now, the only practical way to determine $h_1(x,Q^2)$
was muon pair production (the ``Drell-Yan'' process) with transversely
polarized target and beam.$^{\JaffeJi,\RS}$

In this Letter we show how to
generalize the spin, twist
and chirality analysis of deep inelastic processes to include quark
{\it fragmentation\/} functions.  Our analysis is complete
at the leading order and, for special cases of interest,
at orders $1/\sqrt{Q^2}$ and $1/Q^2$.
As an application of this formalism
-- one of many -- we show how a chirally odd fragmentation function can be
exploited to enable a measurement of $h_1(x,Q^2)$ to be
obtained in polarized
electroproduction of pions from a transversely polarized nucleon.
This is an
experiment which could be performed at several existing facilities.
Related suggestions involving semi--inclusive production of
$\Lambda$--hyperons and of two pions have been discussed
previously.$^{\Collins,\Artru}$~ Our proposal is simpler since it
involves only one particle in the final state and does not require
measurement of that particle's spin.   The price we pay for this
simplicity is suppression by a power of $\sqrt{Q^2}$.

The simplest quark fragmentation function
is represented diagramatically in Fig.~[\fragfunction].  More complicated
fragmentation processes, such as coherent fragmentation of several quarks
and gluons, do contribute at order
$1/\sqrt{Q^2}$ and beyond.  For reasons discussed below, they will not
concern us here.  In Fig.~[\fragfunction], a quark of momentum $k$ and
helicity $h$ fragments into a hadron of momentum $P$ and helicity $H$ plus
an unobserved final state $X$.  The process then repeats in reverse as the
unobserved system, $X$, plus the hadron of momentum $P$ and helicity $H'$
reconstitute the quark of momentum $k$ and helicity $h'$.  The scattering
$k+P\to k+P$ is forward, {\it i.e.\/} collinear.  For definiteness, we take
the momentum of the quark--hadron system to be aligned along the $\hat
e_3$--axis.  Then helicity is conserved as a consequence of angular
momentum conservation about this axis:  $h-H=h'-H'$.  The initial and final
hadron helicities $H$ and $H'$ need not be equal because the hadron need
not have been in a helicity eigenstate; likewise for the quark.  This
possibility arises when observed hadrons are polarized transversely to the
direction of hard momentum flow in a deep inelastic process.$^{\JaffeJi}$

Our first objective is to classify the spin and chirality dependence
and twist (order in $1/Q^2$ as $Q^2\to\infty$) of the possible quark
fragmentation functions pictured in Fig.~[\fragfunction].
To do this it is necessary
to decompose the Dirac spin space of the quark field component with momentum
along the $\hat e_3$--axis.  Consider the three mutually compatible ({\it
i.e.\/} commuting) sets of projection operators,
$$
\eqalignno{
P_\pm &= {1\over 2}\gamma^{\mp}\gamma^{\pm} = {1\over2} (1\pm\alpha_3)
&\eqname\lc\cr
\Lambda_\pm &= {1\over 2}(1\pm\sigma_3)
&\eqname\helicity\cr
\chi_\pm &= {1\over 2}(1\pm\gamma_5) &\eqname\chirality\cr}
$$
$P_\pm$ projects on the ``good'' and ``bad'' light--cone
components of the quark field, respectively.  $\Lambda_\pm$ and $\chi_\pm$
project on positive and negative helicity and chirality states
respectively.  It is easy to show, then, that the good light--cone
component of the quark field with positive (negative) helicity,
$\psi_{\uparrow +}\equiv \Lambda_+P_+\psi$ ($\psi_{\downarrow +}\equiv
\Lambda_-P_+\psi$), has positive (negative) chirality.  In contrast, the
bad light--cone component with positive (negative) helicity, $\psi_
{\uparrow -}\equiv \Lambda_+P_-\psi$ ($\psi_{\downarrow -}\equiv
\Lambda_-P_-\psi$), has {\it negative\/} ({\it positive\/}) chirality.

Our studies have shown$^{\JaffeJi,\Ji}$ that
\Item{(1)}
Quark fragmentation functions of the form shown in
Fig.~[\fragfunction] and the equivalent gluon fragmentation
functions (without further active parton lines) are sufficient to
characterize hadron production in hard processes, provided:
(i) one studies leading twist (twist-two)
(${\cal O}(1/Q^0)$) in any hard process, or
(ii) one studies an effect in deep inelastic lepton scattering at
the lowest twist at which it arises, and one ignores QCD radiative
corrections.
\Item{(2)} Each appearance of a bad
component of the quark field costs one power of $\sqrt{Q^2}$ in the deep
inelastic limit ({\it i.e.\/} it increases the twist by unity);
\Item{(3)} For produced hadrons of spin-1/2, helicity differences
are observed in longitudinal spin asymmetries; helicity flip is observed
in transverse spin asymmetries;
\Item{(4)} Perturbative QCD cannot flip quark chirality (except
through quark mass insertions which we assume to be negligible for light
quarks) so chirally--odd quark distribution and fragmentation functions
must occur in pairs.

\noindent The first two rules emerge from a detailed study of the
operator product expansion$^{\JaffeSoldate}$ or equivalently
the collinear expansion of Feynman diagrams.$^{\EFP}$~  Rule (1.i) is
well-known and corresponds to the usual probabilistic formulation of the
parton model at twist-two.  Rule (1.ii) is a new result presented in detail
in Ref.~[\Ji].  As examples consider two {\it distribution\/} functions to
which the rule (1.ii) applies:  transverse polarization
($g_2$) or longitudinal ($F_L$) effects in deep inelastic
lepton scattering.  In the absence of QCD radiative corrections, these
effects first appear at twist-three ${\cal O}(1/\sqrt{Q^2})$ and twist-four
${\cal O}(1/Q^2)$ respectively.  There are several multiquark/gluon
distribution functions which cannot be reduced to Fig.~[\fragfunction] which
might be expected enter $g_2$ or $F_L$.  In the case of $g_2$ it is well
known since the work of Shuryak and Vainshteyn$^{\ShuryakVainshteyn}$ that
all contributing operators at twist-three can be arranged
by careful use of the QCD equations of motion in the form of a
quark-quark correlation function evaluated in the target state.
The same result applies to $F_L$, in
this case at twist-four.  This result allows us to use the properties of
two-particle forward amplitudes to catalogue the quark distribution and
fragmentation functions which control hadron production at the leading
non-trivial twist in deep inelastic scattering.
Rule (3)
is a simple consequence of quantum mechanics:  transversely polarized
states are linear combinations of helicity eigenstates.  The final rule is
obvious since QCD and the electroweak interactions are all
chirally invariant in perturbation theory neglecting mass insertions.

We now combine the above classification of quark fields
with these rules to enumerate and characterize
quark fragmentation.
Fragmentation functions can be labelled uniquely by specifying the
helicity of quarks and hadrons and the light cone projection of the quarks
in Fig.~[\fragfunction]:
$\hat A^{ab}_{hH;h'H'}$, where $a$ and $b$ are the quark
light--cone projections, either $+$ or $-$.
Parity invariance of QCD requires:
$\hat A^{ab}_{hH;h'H'}= \hat A^{ab}_{-h-H;-h'-H'}$.
Time reversal
invariance, which further reduces the number of independent quark {\it
distribution functions\/} does not generate relationships among the
$\{\hat A\}$ because it changes the {\it out\/}--state $(PX)_{\rm out}$ in
Fig.~[\fragfunction] to an {\it in\/}--state.

As a first example,
consider production of a scalar meson like the pion.  Through order
$1/\sqrt{Q^2}$
there are three independent fragmentation functions:  $\hat A^{++}_{{1\over
2}0;{1\over 2}0}$, $\hat A^{+-}_{{1\over 2}0;{1\over 2}0}$, and
$\hat A^{-+}_{{1\over 2}0;{1\over 2}0}$.  The first is twist-two and scales
in the $Q^2\to\infty$ limit, the latter two are twist-three and are
suppressed by $1/\sqrt{Q^2}$ in the $Q^2\to\infty$ limit.
The first function, $\hat A^{++}_{{1\over 2}0;{1\over 2}0}$, is
proportional to the traditional fragmentation function $D(z,Q^2)$.  It has
the same twist, light-cone, helicity and chirality structure as the
familiar, spin-average quark distribution function, $f_1(x,Q^2)$, so to
avoid an explosion of notation we denote it by $\hat f_1(z,Q^2)$
[We will follow the same convention for other fragmentation functions.]:
$$
\hat f_1(z,Q^2) \propto \hat
A^{++}_{{1\over 2}0;{1\over 2}0}
\eqname\f
$$
If we were studying quark {\it distribution}
functions, the latter two would be equal by time-reversal invariance.
Here, there are two independent fragmentation functions.
$$
\eqalignno{
\hat e_1(z,Q^2) &\propto \hat A^{+-}_{{1\over2}0;{1\over 2}0}
+ \hat A^{-+}_{{1\over 2}0;{1\over 2}0} \cr
\hat e_{\bar 1}(z,Q^2) &\propto \hat A^{+-}_{{1\over 2}0;{1\over 2}0}
- \hat A^{-+}_{{1\over 2}0;{1\over 2}0} &\eqname\ee\cr}
$$
The application to spin--1/2 is summarized in Table~1.
The fragmentation functions described in Eqs.~\f\ -- \ee\
(for spin-zero) and in Table~1 (for spin $1/2$)
are sufficient to describe quark fragmentation in
processes to which Rule (1.i) or (1.ii) applies.

In order to relate particular deep inelastic processes to quark
distribution and fragmentation functions and to study them in models of
non-perturbative QCD, it is necessary to have operator representations for
them.  We presented this formalism for distribution functions in
Ref.~[\JaffeJi].  Since we are interested in pion production
here, we study fragmentation functions which are independent
of the final hadron's spin.  The generalizions to spin--1/2 and spin--1
are presented in Ref.~[\Ji].  Generalizing the procedure in
Refs.~[\JaffeJi] and [\CollinsSoper], we can define
four fragmentation  functions with quark fields alone,
$$\eqalignno{
z \, \int {d\lambda \over 2\pi}
           e^{-i\lambda/ z}
       \left\langle 0 \right| \gamma^{\mu}\psi(0)
       \left| PX_{\rm out} \right\rangle \!
       \left\langle PX_{\rm out} \right| \bar \psi(\lambda n)
       \left| 0 \right\rangle
          &= 4[\hat f_1(z) p^{\mu} + \hat f_4(z) M^2n^{\mu}],
          &\eqname\fstar\cr
z \, \int {d\lambda \over 2\pi}
           e^{-i\lambda / z}
       \left\langle 0 \right| \psi(0)
       \left| PX_{\rm out} \right\rangle \!
       \left\langle PX_{\rm out} \right| \bar \psi(\lambda n)
       \left| 0 \right\rangle
          &= 4M \hat e_1(z),
          &\eqname\estar\cr
z \, \int {d\lambda \over 2\pi}
           e^{-i\lambda/ z}
       \left\langle 0 \right| \sigma^{\mu\nu}i\gamma_5\psi(0)
       \left| PX_{\rm out} \right\rangle \!
       \left\langle PX_{\rm out} \right| \bar\psi(\lambda n)
       \left| 0 \right\rangle
         &= 4M\epsilon^{\mu\nu\alpha\beta}p_\alpha n_\beta
            \hat e_{\bar 1}(z), &\eqname\eonestar\cr}
$$
where $P$ is the four-momentum of the pion and $p$ and
$n$ are two light-like vectors such that
$p^2=n^2=0$, $p^-=n^+=0$, $p\cdot n =1$, and $P^{\mu}=p^{\mu} +
n^{\mu}m_\pi^2/2$. All Dirac indices on quark fields are implicitly
contracted.
The mass $M$ appearing in Eqs.~\fstar\ -- \eonestar\ is a generic QCD mass
scale, which we sometimes choose for convenience to be the nucleon mass.
We avoid use of the produced hadron mass because of
the singular behavior introduced in the chiral limit (the
left hand side of Eq.~\estar\ or \eonestar\ does not vanish as
$m_\pi\rightarrow 0$). The summation over $X$ is implicit and covers
all possible states which can be populated by the
quark fragmentation. The state $\left| PX_{\rm out} \right\rangle$ is
an out state between the pion and $X$.
The renormalization scale dependence is suppressed
in Eqs.~\fstar\ -- \eonestar.  Here we work in $n\cdot A = 0$
gauge, otherwise gauge links have to be added to ensure
the color gauge invariance.  The gauge invariance
and other issues of interpretation for equations like
Eqs.~\fstar -- \eonestar\
are discussed in detail in Ref.~[\CollinsSoper].
{}From a simple dimensional analysis,  we see that $\hat f_1(z)$, $\hat
e_1(z)$ and  $\hat e_{\bar 1}(z)$, and $\hat f_4(z)$
are twist-two, -three, and -four, respectively (they contribute in order
$1/Q^0$, $1/Q^1$, and $1/Q^2$ respectively);
and from their $\gamma$-matrix structure,
$\hat f_1(z)$ and $\hat f_4(z)$ are chirally even
and $\hat e_1(z)$ and $\hat e_{\bar 1}(z)$ are chirally odd.  This
assignment agrees with the results quoted above.  Hermiticity guarantees
that these fragmentation functions are real.

As an important application of the new fragmentation
functions introduced above, we consider
deep-inelastic scattering with
longitudinally polarized leptons on
polarized nucleon targets, focusing on pion production
in the current fragmentation region. As we shall
show below, this process allows us to gain access to
the nucleon's transversity distribution.

The simplest cut diagram for the process is
shown in Fig.~[\transversity], where a quark struck by
the virtual photon fragments into an observed pion
plus other unobserved hadrons. The cross section
of the process is proportional to a trace and integral over the
quark loop which contains the quark distribution
function and fragmentation function.
Due to chirality conservation at the hard (photon) vertex,
the trace picks up only
the products of the terms in which the distribution and
fragmentation functions have the same chirality (Rule~(4) above).
When the nucleon is longitudinally polarized
(with respect to the virtual-photon momentum),
the twist-two, chirally even distribution $g_1(x)$ can couple
with the twist-two chirally even fragmentation function
$\hat f_1(z)$, producing a leading contribution ${\cal O}(1/Q^0)$
to the cross
section. On the other hand, in the case of a transversely
polarized nucleon, there is no
leading-order contribution.
At the next order, the nucleon's transversity
distribution $h_1(x)$ can combine with the twist-three
chirally odd fragmentation function $\hat e_1(z)$,
and similarly $g_T(x)$ can combine with the chirally even
transverse-spin distribution $\hat f_1(z)$. Both couplings produce
$1/\sqrt{Q^2}$ contributions to the cross section.

It is simple to see, however, that Fig.~[\transversity] alone
does not produce an electromagnetically-gauge-invariant
result.  This is a typical example of the need to consider
multi-quark/gluon processes beyond twist-two.$^{\JaffeSoldate,\EFP}$~
In the present case (twist-three),
however, the contributions from coherent
scattering can be expressed, with novel use of QCD
equations of motion, in terms
of the distributions and fragmentation functions
defined from quark bilinears.  This is a specific example of Rule (1.ii).
The combined result is
gauge invariant, as can be seen from the resulting
nucleon tensor,
$$
\hat W^{\mu\nu} = -i\epsilon^{\mu\nu\alpha\beta}
{q_{\alpha}\over \nu}
[(S\cdot n)p_\beta \hat G_1(x,z) +
{S_{\perp \beta}} \hat G_T(x,z)]
\eqn\tensor
$$
where $S$ is the polarization vector of the nucleon
($S^\mu = (S\cdot n)p^\mu + (S\cdot p)n^\mu + S_\perp^\mu$),
$p$ and $n$ are light-cone
vectors defined with respect to the virtual-photon momentum $q$.
The two structure functions in $\hat W^{\mu\nu}$ are related to
parton distributions and fragmentation functions,
$$
\eqalignno{
\hat G_1(x,z) &= {1\over 2} \sum_a e_a^2 g_1^a(x) \hat f_1^a(z) \cr
\hat G_T(x,z) &= {1\over 2} \sum_a e_a^2 \Big [g^a_T(x) \hat f^a_1(z)
                 + {h^a_1(x) \over x} {\hat e^a(z)\over z} \Big]
                 &\eqname\GGT\cr}
$$
where the summation over $a$ includes quarks and antiquarks
of all flavors.

To isolate the spin-dependent part of the deep-inelastic
cross section we take the difference of cross sections
with left-handed and right-handed leptons, we use
$$
{d^2\Delta \sigma \over dE'd\Omega}
= {\alpha_{\rm em}^2 \over Q^4} {E'\over EM_N}
\Delta \ell^{\mu\nu}\hat W_{\mu\nu}
\eqname\crosssec
$$
where $Q^2 = -q^2$, $k = (E, {\bf k})$ and
$k' = (E', {\bf k'})$ are the incident and outgoing momenta
of the lepton, and $\Delta \ell^{\mu\nu}$
is the spin-dependent part of the lepton tensor,
$\Delta\ell^{\mu\nu} = -{\rm Tr}{[\gamma^{\mu}
\mathrel{\mathop{k\!\!\!/}}'\gamma^{\nu}\gamma_5
\mathrel{\mathop{k\!\!\!/}}]}
= - 4i\epsilon^{\mu\nu\alpha\beta} q_{\alpha}
k_{\beta}$.
It is convenient to express the cross section in terms of
scaling variables in a frame where lepton beam defines the
$\hat e_3$-axis
and the $\hat e_1-\hat e_3$ plane contains the nucleon polarization vector,
which has a polar angle $\alpha$.
In this system, the scattered lepton has polar angles
$(\theta, \phi)$ and therefore the momentum transfer ${\bf q}$
has angles $(\theta, \pi-\phi)$. Then,
$$\eqalignno{
{d^4 \Delta \sigma \over dx\,dy\,dz\,d\phi}
= {8\alpha^2_{\rm em} \over Q^2}
&\Big[ \cos\alpha (1-{y\over 2}) G_1(x,z) \cr
&\quad + \cos\phi\sin\alpha\sqrt{(\kappa -1)(1-y)}
\left(G_T(x,z) - G_1(x,z)(1-{y\over 2})\right)\Big]
&\eqname\final\cr}$$
where $y=1-E'/E$ and $\kappa = 1 + 4x^2M^2/Q^2$ in the second term
signals the suppression by a factor of $1/Q$ associated with
the structure function $G_T$. The existence
of $G_1$ in the same term is due to a small
longitudinal polarization of the nucleon relative to ${\bf q}$
when its spin is perpendicular to the lepton beam.

Eq.~\final\ is our main result. As a check, we multiply
by $z$, integrate over it and sum over all hadron species.
Using the well-known momentum sum rule,
$\sum_{\rm hadrons} \int dz z \hat f_1^a(z) = 1$,
and the sum rule,
$\sum_{\rm hadrons}
\int dz \hat e^a_1(z) = 0$,
which is related to the fact that the chiral condensate
vanishes in the perturbative QCD vacuum,
we get the well known result for {\it total\/} inclusive scattering, given in
Eq.~(2.8) in Ref.~[\Jaffe] (if one neglects the terms of order $1/Q^2$ in the
latter). [Eq (2.8) in Ref.~[\Jaffe] contains a sign error:
the sign of the second term should be reversed
corresponding to the replacement $\cos\phi\to\cos(\pi-\phi)$.]  The
similarity between the inclusive and semi-inclusive cross sections
suggests that they can be extracted conveniently
from the same experiment.

The aim of this example was to show that an unfamiliar fragmentation
function ($\hat e_1$) could be employed to obtain a measurement of an
interesting, if unfamiliar, distribution function ($h_1$).  It is apparent
from Eq.~\final\ that we have been only partially successful:
although the
$h_1^a$ distribution for each quark flavour appears in Eq.~\final,
the sum
over flavors couples it to the unknown flavor dependence of $\hat e_1^a$.
Fortunately, flavor tagging can be used at large--$z$ to identify the
contributions of individual quark flavors.  For $x$ in the valence region
(where one can ignore antiquarks in the nucleon), and $z\to 1$, the dominant
fragmentation, $u\to\pi^+$, $d\to\pi^-$, $s\to K^-$,  effectively allows one
to trigger on the contributions of $u$, $d$ and $s$ quarks separately.  One
might be concerned that the unknown fragmentation function, $\hat e_1$,
might not respect the dominant fragmentation selection rules, which have
only been tested for the spin-averaged, twist-two fragmentation function,
$\hat f_1$.  However, the coherent gluon interactions which distnguish the
twist-three $\hat e_1$ from $\hat f_1$ are flavor independent and should
not alter the selection rules.  More complicated flavor structure does
arise at higher twist where multiquark correlation functions appear.  We
have not attempted to make an estimate of the usefulness of this flavor
tagging method in the manner of Ref.~[\Rith], which should precede the
attempt to carry out this measurement.

\vfill\eject

\hbox to \hsize{\hfil \hbox{\epsffile{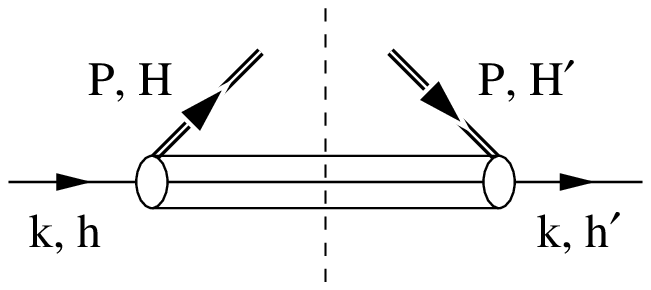}}\hfil}
\bigskip
\centerline{Fig. 1.~~~Diagram for quark fragmentation functions.}

\bigskip\bigskip

\hbox to \hsize{\hfil \hbox{\epsffile{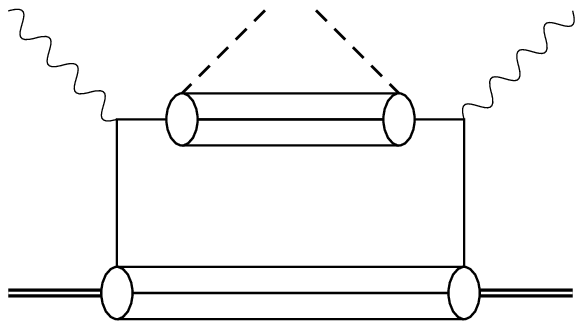}}\hfil}
\bigskip
\centerline{Fig. 2.~~~Pion-production in deep-inelastic scattering.}

\bigskip\bigskip

\midinsert
\bigskip
\baselineskip 15pt
\def\strut{\hbox{\vrule height 4pt depth 4pt width 0pt}}
\centerline{Table I.~~~Quark fragmentation functions for
spin-${1\over2}$ baryon.}
\centerline
{Note: the functions with bar vanish if there are no final state interactions}
$$\vbox{\offinterlineskip
\hrule
\halign{
&\vrule#&\enskip\hfil$#$\hfil\enskip
&\vrule#&\enskip\hfil$#$\hfil\enskip
&\vrule#&\enskip\hfil$#$\hfil\enskip
&\strut#&\enskip\hfil$#$\hfil\enskip
&\vrule#&\enskip\hfil$#$\hfil\enskip\cr
height5pt&&&\omit&&\omit&&\omit&&\omit&\cr
&&&${\rm Twist-2}$&&\multispan3\hfil{\rm Twist-3}\hfil
&&${\rm Twist-4}$&\cr
&&&&&&&&&&\cr
&&&++&&+-${\rm (S)}$&&+-${\rm (A)}$&&--&\cr
height5pt&&&&&&&&&&\cr
\noalign{\hrule}
height4pt&&&\omit&&\omit&&\omit&&\omit&\cr
& \skew6\hat{A}_{{1\over2}{1\over2} \to {1\over2}{1\over2}}
+ \skew6\hat{A}_{{1\over2}{-{1\over2}} \to {1\over2}{-{1\over2}}}
&& \skew4\hat f_1
&& \hat e_1
&& \hat e_{\bar{1}}
&& \skew4\hat f_4
&\cr
height4pt&\omit&&\omit&&\omit&&\omit&&\omit&\cr
\noalign{\hrule}
height4pt&\omit&&\omit&&\omit&&\omit&&\omit&\cr
& \skew6\hat{A}_{{1\over2}{1\over2} \to {1\over2}{1\over2}}
- \skew6\hat{A}_{{1\over2}{-{1\over2}} \to {1\over2}{-{1\over2}}}
&& \hat g_1
&& \hat h_2
&& \hat h_{\bar{2}}
&& \hat g_3
&\cr
height4pt&\omit&&\omit&&\omit&&\omit&&\omit&\cr
\noalign{\hrule}
height4pt&\omit&&\omit&&\omit&&\omit&&\omit&\cr
& \skew6\hat{A}_{{1\over2}{-{1\over2}} \to {-{1\over2}}{1\over2}}
&& \hat h_1
&& \hat g_2
&& \hat g_{\bar{2}}
&& \hat h_3
&\cr
height4pt&\omit&&\omit&&\omit&&\omit&&\omit&\cr
\noalign{\hrule}}}$$
\endinsert

\endpage
\singlespace
\refout
\figout
\end